\documentclass[11pt]{article}
\usepackage[margin=1in]{geometry}
\usepackage[utf8]{inputenc}
\usepackage[T1]{fontenc}

\usepackage{amsmath,amssymb,amsthm,amsfonts}
\usepackage{graphicx}
\usepackage{float}      
\usepackage{subcaption}
\usepackage{arydshln}
\usepackage{authblk}
\usepackage{mathtools}
\usepackage{multirow}

\usepackage{bm}  
\usepackage{mathrsfs}   
\usepackage{setspace}

\usepackage{hyperref}
\hypersetup{
    colorlinks,
    linkcolor={blue!80!black},
    urlcolor={blue!80!black},
    citecolor=teal
}
\usepackage{url}
\usepackage[dvipsnames]{xcolor}
\usepackage[title]{appendix}

\usepackage{tikz}  
\usepackage{tikz-cd}  
\usetikzlibrary{positioning}
\usepackage{tikz-qtree}
\usepackage[all]{xy}
\usepackage{mathdots}       
\usepackage{caption}
\usepackage[capitalize]{cleveref}

\usepackage{thmtools}
\usepackage{thm-restate}

\allowdisplaybreaks

\newtheorem{thm}{Theorem} 
\newtheorem{prop}[thm]{Proposition}
\newtheorem{cor}[thm]{Corollary}
\newtheorem{lem}[thm]{Lemma}
\newtheorem{conj}[thm]{Conjecture}

\theoremstyle{definition}

\newtheorem{rmk}[thm]{Remark}

\newtheorem{construction}[thm]{Construction}

\newtheorem{claim}[thm]{Claim}

\newtheorem*{thm*}{Theorem}
\newtheorem*{lem*}{Lemma}
\newtheorem*{prop*}{Proposition}
\newtheorem*{defn*}{Definition}
\newtheorem*{prob*}{Problem}
\newtheorem*{ques*}{Question}

\DeclareMathOperator{\diag}{diag}

\newcommand{\F}{\mathbb F}

\newcommand{\eps}{\varepsilon}

\newcommand{\nc}{\newcommand}

\newcommand{\mb}{\mathbb}
\newcommand{\mbf}{\mathbf}

\nc{\bco}[1]{\{0,1\}^{#1}}

\usepackage{qcircuit}
\newcommand{\ket}[1]{| #1 \rangle}
\newcommand{\bra}[1]{\langle #1|}

\usepackage{algorithm}
\usepackage{algorithmic}



\newcommand{\be}{\begin{equation}}
\newcommand{\ee}{\end{equation}}
\newcommand{\bes}{\begin{equation*}}
\newcommand{\ees}{\end{equation*}}
\newcommand{\bea}{\begin{eqnarray}}
\newcommand{\eea}{\end{eqnarray}}
\newcommand{\beas}{\begin{eqnarray*}}
\newcommand{\eeas}{\end{eqnarray*}}
\newcommand{\bal}{\begin{aligned}}
\newcommand{\eal}{\end{aligned}}

\usepackage{float}
\usepackage{cite}

\def\1{\mbf{1}}

\def\na{\mathrm{na}}
\def\ad{\mathrm{ad}}

\title{Exponential quantum-classical query separations for cyclic Forrelation}

\author[1]{Yuxin Zhang\thanks{zhang.yuxin@renyi.hu}}

\author[2]{Changpeng Shao\thanks{changpeng.shao@amss.ac.cn}}

\affil[1]{{\small Alfr\'{e}d R\'{e}nyi Institute of Mathematics, Budapest, Hungary}}

\affil[2]{{\small SKLMS, Academy of Mathematics and Systems Science, Chinese Academy of Sciences, Beijing, China}}

\date{\today}

\begin{document}

\maketitle

\begingroup
\hypersetup{citecolor=blue!80!black}

\begin{abstract}

We study the randomized classical query complexity of cyclic Forrelation, a cyclic analogue of the $k$-Forrelation problem introduced by Aaronson and Ambainis~\cite[\textcolor{blue!80!black}{STOC'15}]{aaronson2015forrelation} and known to achieve an optimal separation between quantum and randomized query complexities~\cite[\textcolor{blue!80!black}{STOC'21}]{bansal2021k}. 
In $k$-cyclic Forrelation, we are given \(k\) Boolean functions \(f_1,\ldots,f_k:\mathbb{F}_2^n\to\{-1,1\}\), and the goal is to estimate $T_k(f_1,\ldots,f_k)
=
\frac{1}{N}\operatorname{Tr}
(D_{f_1}H^{\otimes n}  \cdots D_{f_k}H^{\otimes n}),$
where $D_{f_i}=\operatorname{diag} (f_i(x): x\in\mathbb{F}_2^n)$, $H$ is the Hadamard gate, and \(N=2^n\).
The problem arises naturally as a normalized-trace estimation problem in the DQC1 query model and can be solved with $k$ quantum queries. Its classical query complexity was recently posed as an open problem in~\cite[\textcolor{blue!80!black}{FOCS'26}]{ji2026dqc1}, which conjectured a lower bound of $\widetilde{\Omega}(N^{1-2/k})$.

We establish a nearly complete characterization of the nonadaptive classical query complexity of cyclic Forrelation. For even \(k\ge 4\), we prove a tight bound of $\Theta_k (N^{1-2/k} )$ for constant additive error. For odd $k\ge 5$, we obtain a lower bound of \(\Omega_k(N^{1-2/(k-1)})\) and an upper bound of \(O_k(N^{1-2/k})\). Surprisingly, the case \(k=3\) admits a polylogarithmic-query nonadaptive classical algorithm, disproving the previous conjecture in this case. These results suggest an unexpected parity-dependent structure: the nonadaptive classical query complexity may be governed by the independence number of the $k$-cycle, leading to the exponent \(1-1/\lfloor k/2\rfloor\), rather than the previously conjectured \(1-2/k\), which was motivated by a Fourier-growth heuristic. Finally, allowing adaptivity, we prove a lower bound of \(\Omega(N^{1/4})\) for every \(k\ge4\).

Although formally close to $k$-Forrelation, the cyclic structure leads to rather different classical behavior, and the standard techniques for $k$-Forrelation do not apply.
Our lower bounds are instead based on hidden-shift collision constructions, and our upper bounds follow from randomized sampling together with cycle-specific conditional-moment estimates. 
While recent work of Girish~\cite[\textcolor{blue!80!black}{STOC'26}]{girish2026fourier} has revealed strong limitations of the model, our results provide a systematic understanding of its query advantages over classical computation.

\end{abstract}

\endgroup

\newpage 

\section{Introduction}
\label{sec:intro}

Forrelation is a fundamental problem in the study of quantum advantages in the query complexity model. Introduced by Aaronson~\cite{aaronson2010bqp} and generalized by Aaronson and Ambainis~\cite{aaronson2015forrelation},
the $k$-Forrelation problem gives query access to $k$ Boolean functions
$f_1,\ldots,f_k:\mathbb F_2^n\to\{-1,1\}$, and the goal is to estimate
\be
\begin{aligned}
    \Phi_k(f_1,\ldots,f_k)
&=\bra{0^n}H^{\otimes n}D_{f_1}H^{\otimes n}\cdots
D_{f_k}H^{\otimes n}\ket{0^n} \\
&=\frac{1}{N^{(k+1)/2}}
\sum_{x_1,\ldots,x_k\in\mathbb F_2^n}
f_1(x_1)\cdots f_k(x_k)\,
(-1)^{x_1\cdot x_2+\cdots+x_{k-1}\cdot x_k},
\end{aligned}
\label{eq:Forrelation}
\ee
where $N=2^n$, $D_{f_i}:=\diag(f_i(x):x\in \mathbb{F}_2^n)$ and $H$ is the Hadamard gate.
A quantum algorithm can estimate $\Phi_k$ using $\lceil k/2\rceil$ queries~\cite{aaronson2015forrelation}. 
On the classical side, it is known that any $q$-query quantum algorithm can be simulated by a randomized classical algorithm using $O(N^{1-1/2q})$ queries~\cite{aaronson2015forrelation, bravyi2021classical}. 
So for constant additive error, $k$-Forrelation admits a randomized classical algorithm with query complexity $O(N^{1-1/k})$~\cite[Theorem 4]{bravyi2021classical}. 
Following earlier progress by Tal \cite{Tal2020} on the related $k$-Rorrelation problem, Bansal and Sinha~\cite{bansal2021k} established a matching lower bound of $\widetilde\Omega(N^{1-1/k})$, showing that the randomized query complexity of $k$-Forrelation is tight up to logarithmic factors.
Sherstov, Storozhenko, and Wu~\cite{SherstovStorozhenkoWu2021} independently obtained a similar result for $k$-Rorrelation.
Moreover, $k$-Forrelation (with $k$ growing polynomially in $n$) is BQP-complete~\cite{aaronson2015forrelation}, and it underlies the oracle separation of BQP and the polynomial hierarchy~\cite{raz2019oracle}. 
Beyond these results, Forrelation and its variants have also been used to establish exponential quantum-classical separations in communication complexity~\cite{GirishRazTal2021,GirishRazZhan2021} and in the query complexity of matrix functions~\cite{MontanaroShao2024}, to prove further oracle separations in quantum complexity theory~\cite{AaronsonIngramKretschmer2022}, to separate quantum query algorithms with different numbers of rounds of adaptivity~\cite{GirishSinhaTalWu2024}, and to construct quantum cryptographic primitives relative to classical oracles~\cite{KretschmerQianSinhaTal2023}, etc.

In this paper, we study how much quantum query advantage is possible in a restricted
model of quantum computation.
The \emph{one-clean-qubit model}, also known as DQC1 (deterministic quantum computation with one qubit), was introduced by Knill and Laflamme~\cite{knill1998power} to capture the computational power of quantum systems whose state is almost completely mixed. 
In this model, a polynomial-size
unitary circuit acts on the initial state
\[
\rho=\ket{0}\bra{0}\otimes\frac{I_{2^n}}{2^n},
\]
in which only the first qubit is pure and is measured at the end of the computation. Although DQC1 is contained in BQP and is believed to be strictly weaker, it provides a natural setting for understanding the resources underlying quantum computational advantages.

Despite its simple structure and limited purity, DQC1 can efficiently solve certain
problems for which no efficient classical algorithms are known.
A natural task in this model is estimating normalized traces of unitaries.
If one applies a Hadamard gate to
the clean qubit, then a unitary $U$ controlled by it, and then another Hadamard gate, the clean qubit is
measured as $0$ with probability
\[
\frac12+\frac{\operatorname{Re}\operatorname{Tr}(U)}{2^{n+1}},
\]
so $O(\varepsilon^{-2})$ repetitions suffice to estimate
$2^{-n}\operatorname{Re}\operatorname{Tr}(U)$ to additive error
$\varepsilon$. 
This problem is DQC1-complete~\cite{knill1998power,shepherd2006computation}, and underlies many other natural DQC1-complete problems, such as approximating Jones polynomials~\cite{shor2008estimating} and estimating spectral sums such as the normalized trace of the inverse~\cite{edenhofer2025dequantization}; see~\cite{ji2026dqc1} and the references therein. 
Moreover, it was shown that, for suitable variants of the model, efficient classical simulation of DQC1 would collapse the polynomial hierarchy~\cite{morimae2014hardness, fujii2018impossibility}.
These results provide complexity-theoretic evidence for the computational advantage of DQC1 over classical computation. 
A complementary approach is to establish unconditional quantum-classical separations, as has been achieved in the one-clean-qubit communication model \cite{KlauckLim2021,arunachalam2023one}. In this work, we turn to query complexity, which provides another natural framework for studying the computational advantages of DQC1 unconditionally.

The DQC1 query model has the same alternating circuit structure as the standard quantum query model, with the restrictions on initial states and measurement described above. The input consists of $k$ Boolean functions $f_1,\ldots,f_k:\mathbb F_2^n\to\{-1,1\}$, accessed through their phase oracles $\{D_{f_j}\}_{j\in[k]}$. In a $q$-query DQC1 algorithm, the unitary $U$ in the circuit above has the form
\[
U=V_{q+1}O_qV_q\cdots O_1V_1,
\]
where $V_1,\ldots,V_{q+1}$ are unitaries independent of input and each $O_i\in\{D_{f_1},\ldots,D_{f_k}\}$ is one of the given oracles.\footnote{Since $U$ is applied controlled
by the clean qubit, each $O_i$ is implemented as a controlled-$D_{f_j}$,
which counts as one query. Each controlled-\(D_{f_j}\) operation can be implemented using one query to the standard oracle and Hadamard gates on the answer qubit~\cite[Section~2.2]{girish2026fourier}.}
Recently, Girish~\cite{girish2026fourier} showed that $2$-Forrelation requires $N^{\Omega(1)}$ queries in the DQC1 model, establishing an exponential query complexity separation between DQC1 and BQP.

While this result highlights the limitations of DQC1 relative to BQP, a complementary question is whether DQC1 can nevertheless offer substantial query advantages over classical computation. 
To establish such an advantage, we seek problems that DQC1 can solve efficiently but are hard for classical algorithms.
The central role of normalized trace estimation suggests a natural candidate: a cyclic analogue of Forrelation. 
Whereas standard Forrelation in \eqref{eq:Forrelation} is defined by a matrix element of a product of Hadamard transforms and diagonal phase oracles, replacing the matrix element by a normalized trace leads to a cyclic analogue.
This seemingly simple modification makes the problem naturally suited to DQC1, while its classical query complexity remains unclear.
A lower bound on its classical query complexity was conjectured in~\cite{ji2026dqc1}.
In this paper, we systematically study the randomized classical query complexity of this problem, in both the nonadaptive and adaptive settings.

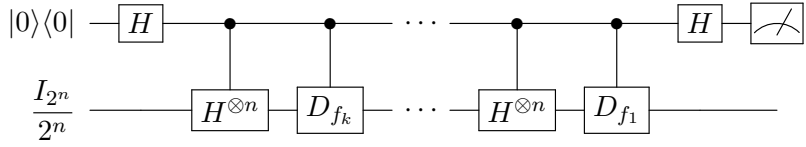
\begin{figure}[ht]
\centering
\[
\Qcircuit @C=1em @R=1.5em {
\lstick{\ket{0}\bra{0}} & \gate{H} & \ctrl{1} & \ctrl{1} & \qw & \cdots & & \ctrl{1} & \ctrl{1} & \gate{H} & \meter \\
\lstick{\displaystyle \frac{I_{2^n}}{2^n}} & \qw & \gate{H^{\otimes n}} & \gate{D_{f_k}} & \qw & \cdots & & \gate{H^{\otimes n}} & \gate{D_{f_1}} & \qw & \qw
}
\]
\caption{A $k$-query DQC1 circuit for $k$-cyclic Forrelation.}
\label{fig:dqc1-cyclic}
\end{figure}

\paragraph{Cyclic Forrelation.}
We consider a cyclic analogue of the $k$-Forrelation problem, replacing a matrix entry by a normalized trace:
\begin{equation}
\label{eq:intro-Tk}
\begin{aligned}
T_k(f_1,\ldots,f_k)
&=\frac1N\operatorname{Tr}\left(
D_{f_1}H^{\otimes n}D_{f_2}H^{\otimes n}\cdots D_{f_k}H^{\otimes n}
\right) \\
&=\frac{1}{N^{k/2+1}}
\sum_{x_1,\ldots,x_k\in\mathbb F_2^n}
f_1(x_1)\cdots f_k(x_k)\,
(-1)^{x_1\cdot x_2+\cdots+x_{k-1}\cdot x_k+x_k\cdot x_1}.
\end{aligned}
\end{equation} 
We define \emph{$k$-cyclic Forrelation} as the problem of estimating $T_k$.
Since all matrices involved are real and orthogonal, $T_k$ is a real
number in $[-1,1]$. 
Compared with $\Phi_k$, the phase differs in a single term: the phase
of $\Phi_k$ follows the path $x_1-x_2-\cdots-x_k$, whereas the
additional term $x_k\cdot x_1$ in $T_k$ closes this path into a cycle.
Each run of the DQC1 circuit with
$U=D_{f_1}H^{\otimes n}\cdots D_{f_k}H^{\otimes n}$ makes one query to
each $f_j$, so $T_k$ can be estimated to constant additive error using $k$ queries in DQC1, as shown in Figure~\ref{fig:dqc1-cyclic}. 
Moreover, just as Forrelation is
BQP-complete, when \(k\) is allowed to grow with the input size, estimating $T_k$ is DQC1-complete~\cite[Proposition~1.3]{ji2026dqc1}. Cyclic Forrelation also plays a central role in establishing exponential classical query lower bounds for estimating normalized traces of functions of sparse Hamiltonians~\cite{ji2026dqc1}.

\paragraph{The classical complexity of cyclic Forrelation.}
Much less is known about the classical query complexity of cyclic Forrelation. 
For $k=4$, Shepherd's hidden-subspace construction~\cite[Proposition~3.3.6]{Shepherd2009} establishes a deterministic classical query lower bound $\Omega(N^{1/4})$ for a closely related normalized-trace problem.
However, the randomized classical query complexity of cyclic Forrelation remained largely unexplored.
For general $k$, it was conjectured in~\cite[Conjecture~1.4]{ji2026dqc1} that estimating $T_k$ to constant additive error requires $\widetilde\Omega(N^{1-2/k})$ classical queries. 
Using a similar technique as in~\cite{bansal2021k}, one can prove a lower bound of $\widetilde{\Omega}\left(N^{1-2 / k}\right)$, but only for additive error of order $1/N$, which is too small to be meaningful.
The classical query complexity of cyclic Forrelation thus remained open, even
for $k=4$, a gap remained between the known
$\widetilde\Omega(N^{1/4})$ lower bound and the conjectured
$\widetilde\Omega(N^{1/2})$.

\subsection{Main results}
\label{subsec:main-results}

For $\varepsilon>0$, let $R^{\na}_\varepsilon(T_k)$ and $R^{\ad}_\varepsilon(T_k)$ denote the minimum numbers of queries required by {\em nonadaptive} and {\em adaptive} randomized classical algorithms, respectively, to estimate $T_k(f_1,\ldots,f_k)$ on every input $(f_1,\ldots,f_k)$ to within additive error $\varepsilon$ with probability at least $2/3$.\footnote{We count the total number of queries to all \(k\) functions in the worst case over inputs and internal randomness. The success probability is over the algorithm’s internal randomness. A nonadaptive algorithm fixes all queries before any answers are observed, while an adaptive algorithm may choose queries based on previous answers.}
Our main contribution is a nearly complete picture of the nonadaptive classical query complexity of cyclic Forrelation. Our results are summarized below,
and Table~\ref{tab:query-complexity-summary} gives the bounds for constant error.\footnote{In contrast to Forrelation, when $k=2$, we have $T_2=\widehat{f}_1(0)\widehat{f}_2(0)$, the product of the zeroth Fourier coefficients of $f_1$ and $f_2$, which can be efficiently estimated classically. }

\begin{itemize}
\item Even case (Theorems~\ref{thm:even-lower-bound} and~\ref{thm:even-upper}): For $k=2m\ge 4$ and $0<\varepsilon<1$,
\be
R^{\na}_\varepsilon(T_k)
=\Theta\left(m\min\left\{N,\;N^{1-1/m}\varepsilon^{-2/m}\right\}\right).
\ee
This confirms Conjecture~1.4 of~\cite{ji2026dqc1} for
nonadaptive algorithms. 
 
\item $k=3$ (Proposition~\ref{prop:three-upper}): Surprisingly, $T_3$ can be efficiently estimated to additive
error $\varepsilon$ using $O\big(n\varepsilon^{-4}\log(n/\varepsilon)\big)$ nonadaptive
queries.
This refutes the conjecture for $k=3$, where it predicts $\widetilde\Omega(N^{1/3})$, and contrasts sharply with
$3$-Forrelation, which requires $\widetilde\Omega(N^{2/3})$ randomized
queries~\cite{bansal2021k}.
 
\item Odd case (Theorem~\ref{thm:odd-lower} and Corollary~\ref{cor:odd-upper}): For odd $k\ge5$ and constant $0<\varepsilon<1/2$,
\be
\Omega_k\big(N^{1-2/(k-1)}\big)
\le R^{\na}_\varepsilon(T_k)
\le O_k\big(N^{1-2/k}\big)
\ee
The full $\varepsilon$-dependent bounds match when $\varepsilon\le N^{-1/(k+1)}$
(see Remark~\ref{rmk:odd-explicit}). 
 
\item Adaptive lower bound (Theorem~\ref{thm:adaptive-lower-bound}): 
By adapting Shepherd's hidden-subspace construction \cite{Shepherd2009} to Boolean phase oracles, we establish an $\Omega(N^{1/4})$ lower bound against randomized adaptive classical algorithms for every $k\ge4$.
\end{itemize}

\begin{table}[ht!]
\centering
\renewcommand{\arraystretch}{1.3}
\begin{tabular}{|c|c|c|c|}
\hline
\(k\) &
Nonadaptive lower bound &
Nonadaptive upper bound &
Adaptive lower bound
\\
\hline
\(2\)
&
---
&
\(O(1)\)
&
---
\\

\(3\)
&
---
&
\(O(\log N \log\log N)\)
\,\hyperref[prop:three-upper]{Prop.~\ref*{prop:three-upper}}
&
---
\\

even \(k\ge 4\)
&
\(\Omega_k(N^{1-2/k})\) 
\enspace \hyperref[cor:even-lower-bound-constant]{Cor.~\ref*{cor:even-lower-bound-constant}}
&
\(O_k(N^{1-2/k})\)
\enspace \hyperref[thm:even-upper]{Thm.~\ref*{thm:even-upper}}
&
\(\Omega(N^{1/4})\)
\enspace \hyperref[thm:adaptive-lower-bound]{Thm.~\ref*{thm:adaptive-lower-bound}}
\\

odd \(k\ge 5\)
&
\(\Omega_k(N^{1-2/(k-1)})\)
\,\hyperref[cor:odd-lower-bound-constant]{Cor.~\ref*{cor:odd-lower-bound-constant}}
&
\(O_k(N^{1-2/k})\)
\enspace \hyperref[cor:odd-upper]{Cor.~\ref*{cor:odd-upper}}
&
\(\Omega(N^{1/4})\)
\enspace \hyperref[thm:adaptive-lower-bound]{Thm.~\ref*{thm:adaptive-lower-bound}}
\\
\hline
\end{tabular}
\caption{Classical query complexity of cyclic Forrelation for constant additive error.}
\label{tab:query-complexity-summary}
\end{table}

Our results suggest that the nonadaptive query complexities of \(T_{2m}\) and \(T_{2m+1}\) may have the same polynomial dependence on \(N\), revealing an unexpected parity-dependent structure. In particular, we conjecture that the complexity exponent is determined by \(\lfloor k/2\rfloor\), which is the independence number of the $k$-cycle  for \(k\ge3\), rather than by $k$ itself.

\begin{conj}
\label{conj:parity}
For every \(k\ge2\) and every constant \(0<\varepsilon<1/2\),
\be
R^{\mathrm{na}}_\varepsilon(T_k)
=
\widetilde{\Theta}_k
\big(N^{1-1/\lfloor k/2\rfloor}\big).
\ee
\end{conj}
Our results confirm this conjecture for even \(k\) and, up to polylogarithmic factors, for \(k=3\). For odd \(k\ge5\), our lower bound has the conjectured exponent, while a matching upper bound remains open (see Remark~\ref{rem:parity effect}). For odd \(k\), this prediction differs from the exponent \(1-2/k\) suggested by the Fourier-growth heuristic in~\cite{ji2026dqc1}.
This contrasts with the standard $k$-Forrelation problem, for which Fourier-analytic methods yield the optimal randomized query complexity exponent $1-1/k$ (see, e.g., \cite{bansal2021k}). Our results suggest that the cyclic structure introduces additional features that are not reflected in the Fourier degree or the levelwise $\ell_1$-norms of the Fourier coefficients alone.

\subsection{Technical overview}

\paragraph{Nonadaptive lower bounds.}
We illustrate the main idea behind our lower bound for the case $k=4$.
In this case, $T_4$ admits the following simple expression:\footnote{For $s\in \mathbb{F}_2^n$, define the character $\chi_s(x) = (-1)^{s\cdot x}$. For a function $g:\mathbb F_2^n\to\mathbb C$, we use the normalized Fourier transform $\widehat g(s)=\mathbb E_x[g(x)\chi_s(x)]$, where \(x\) is uniform on \(\mathbb F_2^n\). For background on Fourier analysis of Boolean functions, we refer to \cite{ODonnell2014}.}
\be
\begin{aligned}
T_4
&=\frac{1}{N}\operatorname{Tr}
\left(D_{f_1}H^{\otimes n} D_{f_2}H^{\otimes n} D_{f_3}H^{\otimes n} D_{f_4}H^{\otimes n}\right)\\
&=\sum_{s\in\mathbb{F}_2^n}
\widehat{f_1}(s)\widehat{f_3}(s)
\mathbb{E}_x[f_2(x)f_4(x+s)].
\end{aligned}
\label{eq: Fourier expression of T4}
\ee
If we choose $f_1=f_3=\chi_t$ for some unknown
$t\in\mathbb{F}_2^n$, then
\bes
T_4=\mathbb{E}_x[f_2(x)f_4(x+t)].
\ees
This expression measures the correlation between $f_2$
and a shift of $f_4$. In particular, $T_4=1$ if and only if
$f_2(x)=f_4(x+t)$ for every $x$, while $T_4=-1$ if and only if
$f_2(x)=-f_4(x+t)$ for every $x$.
This observation motivates the following construction.
Let $\sigma\in\{1,-1\}$ be a hidden sign, and let
$A:\mathbb{F}_2^n\to\{1,-1\}$ be a uniformly random Boolean function.
Define
$f_2(x)=A(x), 
f_4(x)=\sigma A(x+t).$
Then we have
$T_4
=\mathbb{E}_x[A(x)\sigma A(x+t+t)]
=\sigma.$

We therefore define our hard input distribution as follows.
Choose $t\in\mathbb{F}_2^n$ and
$\sigma\in\{1,-1\}$ independently and uniformly at random,
and choose an independent uniformly random Boolean function
$A:\mathbb{F}_2^n\to\{1,-1\}$. Set
\bes
\begin{aligned}
f_1(x)&=\chi_t(x),&
f_2(x)&=A(x),\\
f_3(x)&=\chi_t(x),&
f_4(x)&=\sigma A(x+t).
\end{aligned}
\ees
By construction, $T_4=\sigma$ for every input in the support of this distribution.
Thus, estimating \(T_4\) to additive error \(\varepsilon\) suffices to recover \(\sigma\) by the sign of the estimate.

The key observation is that a classical algorithm can
obtain information about $\sigma$ only by finding a collision
between queries to $f_2$ and $f_4$.
Indeed, for any $x,y\in\mathbb{F}_2^n$ satisfying $x=y+t$,
we have
$f_2(x)f_4(y)
=A(x)\sigma A(y+t)
=\sigma.$
Conversely, if no such collision occurs, the sets of entries of \(A\) accessed through \(f_2\) and \(f_4\) are disjoint. Consequently, the joint distribution of the
query answers is independent of $\sigma$, and the algorithm
cannot distinguish $\sigma=1$ from $\sigma=-1$ with
nonzero advantage.

Now consider a randomized nonadaptive algorithm making at most
$T$ queries. Condition on its internal randomness, and let
$Q_2,Q_4\subseteq\mathbb{F}_2^n$ denote the resulting fixed
query sets for $f_2$ and $f_4$, respectively.
Define the collision event
\bes
C=\left\{
\exists (x,y)\in Q_2\times Q_4:
x=y+t
\right\}.
\ees
Since $t$ is uniformly random, for every fixed pair
$(x,y)\in Q_2\times Q_4$, we have $\Pr_t[x=y+t]={1}/{N}.$
By the union bound,
\bes
\Pr_t[C\mid Q_2,Q_4]
\leq \frac{|Q_2||Q_4|}{N}
\leq \frac{T^2}{4N},
\ees
where the last inequality follows from
$|Q_2|+|Q_4|\leq T$ and the AM--GM inequality.
Averaging over the algorithm's internal randomness gives
$\Pr[C]\le {T^2}/{4N}.$
If $C$ does not occur, the query transcript is independent
of $\sigma$. Since $\sigma$ is uniformly random, the algorithm
can correctly determine $\sigma$ with probability at most
$1/2$ in this case. On the other hand, if $C$ occurs,
we upper-bound the success probability by $1$.
Therefore, by conditioning on whether a collision occurs,
\bes
\begin{aligned}
\Pr[\mathrm{success}]
&=\Pr[\mathrm{success}\mid C] \cdot \Pr[C]
  +\Pr[\mathrm{success}\mid\neg C] \cdot  \Pr[\neg C]\\
&\leq \Pr[C]+\frac12\Pr[\neg C]\\
&=\frac12+\frac12\Pr[C]\\
&\leq \frac12+\frac{T^2}{8N}.
\end{aligned}
\ees
Consequently, achieving a constant distinguishing advantage requires $T=\Omega(\sqrt{N}).$

The same idea naturally extends to the general $2m$-function case. Instead of hiding a single shift between two random functions, we introduce $m-1$ independent hidden shifts and random functions, and construct the final function so that all these random functions cancel when the corresponding query locations are correctly matched. We then choose the remaining character functions to ensure that $T_{2m}=\sigma$. To recover $\sigma$, a nonadaptive classical algorithm must find an $m$-way collision satisfying all $m-1$ hidden-shift relations simultaneously. Since each fixed candidate tuple satisfies these relations with probability $1/N^{m-1}$, the probability of finding such a collision with $T$ queries is at most $O(T^m/N^{m-1})$. Consequently, achieving a constant distinguishing advantage requires $T=\Omega(N^{1-1/m})$.

The lower bound for the odd case follows naturally from the even case. 
For odd \(k=2m+1\), the key idea is to modify the even-cycle hard instances using fixed quadratic phases and append one additional known oracle.
This construction transforms the even-cycle identity $T_{2m}=\sigma$ into $T_{2m+1}=\sigma c_n$, where $c_n$ is a known constant satisfying $|c_n|\ge 1/2$. Since the additional phase functions are completely known, every query to the odd-cycle oracles can be simulated using the original even-cycle oracles without increasing the number of queries, while preserving nonadaptivity. Consequently, the odd-cycle construction inherits the same collision-based lower bound as the even case, giving $\Omega(N^{1-1/m})$ nonadaptive classical queries for $2m+1$ functions.

\paragraph{Nonadaptive classical algorithms.}
We again illustrate the main idea for \(k=4\). Write
\[
h(x_1,x_2,x_3,x_4)
=
\left(\prod_{j=1}^4 f_j(x_j)\right) 
(-1)^{(x_1+x_3)\cdot(x_2+x_4)},
\]
so that \(T_4=N\mathbb E[h(X_1,X_2,X_3,X_4)]\) for independent uniform \(X_1,X_2,X_3,X_4 \in \mb F_2^n\).
For each \(j\), we independently sample \(q\) uniform points \(X_{j,1},\ldots,X_{j,q}\in \mb F_2^n\) and query \(f_j\) at these points. We then form the unbiased estimator
\[
\widetilde T_4
=
\frac{N}{q^4}
\sum_{i_1,i_2,i_3,i_4=1}^{q}
h(X_{1,i_1},X_{2,i_2},X_{3,i_3},X_{4,i_4}).
\]
Although the estimator contains $q^4$ terms, it requires only $4q$ oracle queries. Clearly, $\mathbb E[\widetilde T_4]=T_4$. 
The main challenge is to bound the variance, namely, to control the correlations between these terms in $\widetilde T_4$. 
We bound the variance by grouping pairs of summands according to their shared sample indices, as in the standard variance analysis of $U$-statistics~\cite{Hoeffding1948}. Pairs with no shared samples are independent. Among pairs sharing two coordinates, opposite positions in the cycle give the largest contribution. For example, averaging $h$ over $x_2,x_4$ gives
\[
h_{\{1,3\}}(x_1,x_3)
=
f_1(x_1)f_3(x_3)
\widehat f_2(x_1+x_3)\widehat f_4(x_1+x_3).
\]
By Parseval’s identity, its squared \(L_2\) norm is at most \(1/N\), giving a variance contribution of at most \(O(N/q^2)\) as there are $O(q^6)$ pairs of sampling indices with this coincidence pattern. Accounting for all overlap patterns yields
\[
\operatorname{Var}(\widetilde T_4)
=
O\!\left(
\frac1q+\frac{N}{q^2}+\frac{N^2}{q^4}
\right).
\]
Thus \(q\approx \sqrt N\) suffices to estimate \(T_4\) to any prescribed constant additive error with success probability at least \(2/3\).

The same estimator extends to general \(k\). The overlap analysis reflects the geometry of the cycle: shared coordinates containing adjacent positions admit stronger moment bounds than those forming an independent set. For \(1\le q\le N\), the resulting variance bound is
\[
\operatorname{Var}(\widetilde T_k)
=
O_k\!\left(
\frac{N^{r-1}}{q^r}
+
\frac{N^{k-2}}{q^k}
\right),
\qquad r=\lfloor k/2\rfloor.
\]
The two terms arise from maximum independent sets and full overlap, respectively. For even \(k\), both are controlled at the scale \(q=N^{1-2/k}\), matching our constant-error lower bound. For odd \(k\), the independent-set term is controlled at the smaller scale \(N^{1-2/(k-1)}\), but the full-overlap term still requires \(q\approx N^{1-2/k}\). This explains why this estimator does not close the gap in the odd case.

The case $k=3$ admits a substantially faster algorithm based on a different idea. The key observation is that we can express $T_3$ as
\be
T_3=e^{-i\pi n/4}
\sum_{s\in \mathbb{F}_2^n }
i^{\mathrm{wt}(s)}
\widehat F_1(s) \widehat F_2(s) \widehat F_3(s),
\ee
where $\mathrm{wt}(s)$ refers to the Hamming weight of $s$, and $F_j(x)=f_j(x)i^{\mathrm{wt}(x)}$.
By the Parseval identity and the Cauchy–Schwarz inequality, the terms for which $|\widehat F_1(s)|$ is small contribute negligibly to the sum. Thus, it suffices to identify the heavy Fourier coefficients of $F_1$ and estimate the corresponding coefficients of $F_2$ and $F_3$. Using a nonadaptive implementation of the Goldreich–Levin algorithm \cite{goldreich1989hard,ODonnell2014}, we obtain an $O(\log N\log\log N)$-query classical algorithm for constant additive error.

\paragraph{Adaptive lower bounds.}

The nonadaptive hard instances above are easy for adaptive algorithms. For \(k=4\), querying \(f_1=\chi_t\) at the \(n\) standard basis vectors reveals \(t\), after which two further queries determine $\sigma=f_2(t)f_4(0).$
Thus these instances can be solved with only \(n+2\) adaptive queries.

To obtain an adaptive lower bound, we hide this structure using random signs on cosets of a hidden subspace and its orthogonal complement, following Shepherd~\cite{Shepherd2009}. Choose a uniformly random subspace \(L\le\mathbb F_2^n\) of dimension \(\lfloor n/2\rfloor\), independent uniform shifts \(s,t\in\mathbb F_2^n\), and an independent uniform sign \(\sigma\). Let \(\phi\) and \(\psi\) be independent uniformly random sign functions on \(\mathbb F_2^n/L\) and \(\mathbb F_2^n/L^\perp\), respectively. Define
\[
\begin{aligned}
f_1(x)&=\chi_s(x)\phi(x+L),&
f_3(x)&=\chi_s(x)\phi(x+t+L),\\
f_2(x)&=\chi_t(x)\psi(x+L^\perp),&
f_4(x)&=\sigma\chi_s(t)\chi_t(x)\psi(x+s+L^\perp).
\end{aligned}
\]
The orthogonality of \(L\) and \(L^\perp\) ensures that \(T_4=\sigma\), while the random coset signs conceal the character phases. Every query that accesses a previously unused entry of \(\phi\) or \(\psi\) returns a fresh uniformly random sign. Consequently, information about \(\sigma\) can arise only when distinct oracle-input pairs access the same random mask.

The main issue is to control this collision event despite adaptivity. We first run the algorithm against four independent uniformly random Boolean functions. Although the queries depend on previous answers, this ideal query sequence is independent of the hidden subspace and shifts. We couple the real and ideal executions so that they agree until the first repeated mask. Conditioned on the ideal query sequence, any pair of distinct queries accesses the same mask with probability \(O(N^{-1/2})\), by the randomness of the subspace and shifts. A union bound therefore bounds the collision probability after \(q\) queries by
\[
O\!\left(\frac{q^2}{\sqrt N}\right).
\]
Until this event occurs, the transcript carries no information about \(\sigma\), yielding the adaptive lower bound \(\Omega(N^{1/4})\). Padding with constant oracles extends the bound to larger even \(k\), and the quadratic-phase construction extends it to odd \(k\ge5\).

\subsection{Discussion}

We discuss several observations and questions that emerged during the course of our investigation.

\paragraph{The conjectured parity effect.}
Two features of our proofs motivate Conjecture \ref{conj:parity}. On the lower-bound side, for \(k\ge4\), our hard distributions require an \(m\)-way collision, where \(m=\lfloor k/2\rfloor\), yielding the exponent \(1-1/m\). On the upper-bound side, the conditional second-moment bound improves by a factor of \(N\) whenever the overlap pattern contains two adjacent positions of the cycle (Claim~\ref{claim:even-conditional-moments}). Among independent overlap sets, the largest have size \(m\), again giving the query scale \(N^{1-1/m}\).

For odd \(k\), all proper-overlap contributions can be controlled at this conjectured scale. The obstruction in our variance analysis is the full-overlap term, corresponding to two summands using the same samples at every oracle position. This term is of order \(N^{k-2}/q^k\), where \(q\) is the number of samples per oracle, and leads to the larger scale \(N^{1-2/k}\) (Remark~\ref{rem:parity effect}). This restriction arises from the current estimator and does not establish a barrier for other nonadaptive algorithms.

The case \(k=3\) provides evidence that this obstruction can be overcome. The same sampling analysis gives a query bound of \(O(N^{1/3})\), whereas our specialized algorithm uses only polylogarithmically many queries, consistent with the conjectured exponent. That algorithm relies on a Fourier identity specific to the three-cycle. Whether a different estimator or a related algebraic approach can achieve the conjectured exponent for larger odd cycles remains open.

\paragraph{Adaptive versus nonadaptive lower bounds.}

The distinction between adaptive and nonadaptive algorithms is
essential to our lower-bound constructions.
Our nonadaptive hard distribution hides a sign in correlations
between randomly shifted functions, while the shifts themselves
are encoded by character oracles.
An adaptive algorithm can first recover these shifts by querying
the character oracles on the standard basis vectors, and then
query the random-mask oracles at the corresponding shifted locations
to determine the hidden sign.
Consequently, increasing the number of hidden shifts strengthens
the nonadaptive lower bound but does not make this construction hard for adaptive algorithms.
Our adaptive lower bound therefore uses a different construction, in which random functions on hidden cosets conceal the locations of the relevant correlations.

The exponent $1/4$ for adaptive lower bound arises from our coupling argument, which bounds the probability of the first collision between hidden coset labels. A natural approach to strengthening the lower bound for general $k$ would be to introduce more hidden shifts and random coset functions, as in the nonadaptive case. However, this does not immediately lead to stronger adaptive lower bounds. In our construction, the hidden quotient spaces contain approximately $\sqrt N$ cosets, so pairwise collisions already occur at the birthday-paradox scale $q\asymp N^{1/4}$, as in related hidden-subgroup problems~\cite{Simon1997,Shepherd2009,YeHuangLiWang2021}. Thus, our coupling argument, which relies on the absence of such collisions, cannot yield a stronger lower bound.

More importantly, unlike nonadaptive algorithms, adaptive algorithms may exploit these early collisions to learn information about the hidden subspaces or shifts, and then use this information to choose subsequent query locations. For instance, in natural extensions that reuse the same hidden subspace, learning this subspace can make the remaining shifts much easier to recover. Consequently, introducing additional hidden shifts does not necessarily force adaptive algorithms to discover higher-order collisions by chance. Stronger lower bounds may therefore require a more refined analysis of the information revealed by partial collisions, or fundamentally different hard distributions in which the relevant correlations remain difficult to locate even after some hidden structure has been learned. In particular, the exponent $1/4$ is a limitation of our current approach, rather than a known barrier for adaptive query complexity.

\section{Nonadaptive classical query complexity lower bounds}
\label{sec:lower bound}

\subsection{Even-cycle case}
\label{subsec:even-lower}

\begin{construction}[The multi-shift distribution]
\label{construction:even-hard}
Choose independent uniformly distributed shifts $t_1,\ldots,t_{m-1}\in \mathbb{F}_2^n$, and set $t_0=t_m=0$. Let $a_j=t_{j-1}+t_j$ for $j=1,\ldots,m$.
Choose independent uniform random Boolean functions $A_1,\ldots,A_{m-1}: \mathbb{F}_2^n \to\{-1,1\}$, and an independent uniform sign $\sigma\in\{-1,1\}$.
Define
\[
F(x)
=
\sigma\prod_{j=1}^{m-1}A_j(x+t_j).
\]
The $2m$-input functions are
\begin{equation}
\label{eq:even-hard-oracles}
\begin{aligned}
f_{2j-1}(x)&=\chi_{a_j}(x),
&&1\le j\le m,\\
f_{2j}(x)&=A_j(x),
&&1\le j\le m-1,\\
f_{2m}(x)&=F(x).
\end{aligned}
\end{equation}
\end{construction}

The following identity is the algebraic core of the construction.

\begin{prop} 
\label{prop:even-cancellation}
For every realization of Construction
\ref{construction:even-hard}, we have
\be
T_{2m}(f_1,\ldots,f_{2m})=\sigma,
\ee
and
\begin{equation}
\label{eq:even-word-identity}
W:=D_{f_1} H^{\otimes n} D_{f_2} H^{\otimes n} \cdots D_{f_{2m}} H^{\otimes n}
=
\sigma I.
\end{equation}
\end{prop}

\begin{proof}
We first derive a Fourier representation of
$T_{2m}$. Expanding the normalized trace
and grouping the variables at odd and even
positions gives
\beas
T_{2m}
= \frac{1}{N^{m+1}}
\sum_{x_1,\ldots,x_m\in \mathbb{F}_2^n }
\prod_{j=1}^m f_{2j}(x_j) 
\prod_{j=1}^m
\left(
\sum_{y_j\in \mathbb{F}_2^n }
f_{2j-1}(y_j)
(-1)^{y_j\cdot(x_{j-1}+x_j)}
\right),
\eeas
where $x_0=x_m$. By the definition of the
Fourier transform, this becomes
\[
T_{2m}=\frac1N
\sum_{x_1,\ldots,x_m\in \mathbb{F}_2^n}
\prod_{j=1}^m
f_{2j}(x_j)
\widehat f_{2j-1}(x_{j-1}+x_j).
\]
Setting $x=x_m$ and $s_j=x_j+x_m$ for
$1\le j<m$, with $s_0=s_m=0$, we obtain
\beas
T_{2m}
=\sum_{s_1,\ldots,s_{m-1}\in \mathbb{F}_2^n}
\left(
\prod_{j=1}^m
\widehat f_{2j-1}(s_{j-1}+s_j)
\right) 
\mathbb E_x\left[
f_{2m}(x)
\prod_{j=1}^{m-1}f_{2j}(x+s_j)
\right].
\eeas

Now consider the functions in Construction \ref{construction:even-hard}.
Since $f_{2j-1}(x)=\chi_{t_{j-1}+t_j}(x)$ and
$\widehat{\chi_a}(s)=1$ if $s=a$ and 0 otherwise,
the product of Fourier coefficients is
nonzero only when
$
s_{j-1}+s_j=t_{j-1}+t_j,
$ for $j=1,\ldots,m.$
Since $s_0=t_0=0$, these conditions
successively imply
$s_j=t_j$ for all $1\le j<m$,
and the last condition is automatically
satisfied. Thus the Fourier sum reduces
to a single term:
\[
T_{2m}
=\mathbb E_x\left[
f_{2m}(x)
\prod_{j=1}^{m-1}f_{2j}(x+t_j)
\right].
\]
Substituting
$f_{2j}(x)=A_j(x)$ and
$f_{2m}(x)=\sigma\prod_{j=1}^{m-1}A_j(x+t_j)$,
we obtain
\[
T_{2m}
=\sigma\mathbb E_x
\left[\prod_{j=1}^{m-1}A_j(x+t_j)^2\right]
=\sigma.
\]
Finally, since $W$ is unitary and
$\operatorname{Tr}(W)=\sigma N$, we must have $W=\sigma I$, as claimed.
\end{proof}

The Fourier representation also explains the motivation behind Construction \ref{construction:even-hard}. By choosing the odd-indexed oracles as character functions, their Fourier coefficients force the summation variables to equal the prescribed hidden shifts \(t_1,\ldots,t_{m-1}\). The final oracle is then chosen so that all random functions cancel in the resulting correlation, leaving only the hidden sign \(\sigma\). This structure also motivates the \(m\)-way collision argument used in the following lower-bound proof.

\begin{thm}[Even-cycle nonadaptive lower bound]
\label{thm:even-lower-bound}
For every integer $m \ge 2$ and $0<\varepsilon<1$, we have
\begin{equation}
R_{\varepsilon}^{\na}(T_{2m})
=\Omega\left(m \min \big\{N, N^{1-1 / m} \varepsilon^{-2 / m}\big\}\right).
\end{equation}
\end{thm}

\begin{proof}
We use Construction~\ref{construction:even-hard}, but replace the last oracle by
\[
f_{2m}(y)=B(y)\prod_{j=1}^{m-1}A_j(y+t_j),
\]
where $B:\mathbb{F}_2^n\to\{-1,1\}$ is a random sign function, independent of $t_j$ and $A_j$.
The same calculation as in
Proposition~\ref{prop:even-cancellation} gives
$W=H^{\otimes n} D_B H^{\otimes n}.$
Consequently,
\begin{equation}
\label{eq:even-mean-gadget}
T_{2m}
=\frac1N\operatorname{Tr}(H^{\otimes n} D_B H^{\otimes n})
=\frac1N\sum_{y\in\mathbb{F}_2^n}B(y).
\end{equation}

Consider a randomized nonadaptive algorithm making at most
$T$ queries. Condition on its internal randomness, and let
$Q_{2j}\subseteq \mathbb{F}_2^n$ be its fixed query set for $f_{2j}$,
where $1\le j\le m$.
Define the collision event
\[
C=\left\{\,\exists
(x_1,\ldots,x_{m-1},y)
\in Q_2\times\cdots\times Q_{2m}:
x_j=y+t_j,\quad 1\le j\le m-1
\right\},
\]
and let $M=\left|\left\{
y\in Q_{2m}:
y+t_j\in Q_{2j}\text{ for every }1\le j<m
\right\}\right|$, so event $C$ occurs if and only if $M\ge 1$.

We next bound the expected number of collisions.
For any fixed tuple $(x_1,\ldots,x_{m-1},y)$, the collision conditions determine the hidden shifts: $t_j=x_j+y,$ for $1\le j\le m-1.$
Since the shifts are independent and uniformly random,
these conditions hold with probability $N^{-(m-1)}$.
Since $\sum_{j=1}^{m}|Q_{2j}|\leq T$, linearity of expectation and AM--GM inequality give
\[
\mathbb E[M\mid Q_2,\ldots,Q_{2m}]
=
\frac{\prod_{j=1}^{m}|Q_{2j}|}{N^{m-1}}
\le
\frac{(T/m)^m}{N^{m-1}}.
\]
Averaging over the algorithm's internal randomness gives
\begin{equation}
\label{eq:even-revealing-count}
\Pr[C]\le\mathbb E[M]\le\frac{(T/m)^m}{N^{m-1}}.
\end{equation}

First, take $B\equiv\sigma$ so that $T_{2m}=\sigma$ under our hard distribution.
Any $\varepsilon$-additive estimator determines $\sigma$ by taking the sign of its estimate, since $\varepsilon<1$.
When $C$ does not occur, the query answers reveal no information
about the uniform sign $\sigma$.\footnote{When $C$ does not occur, the joint distribution
of all query answers is independent of $B$.
Indeed, for every $y\in Q_{2m}$, at least one factor $A_j(y+t_j)$ has not been directly queried and remains a uniform random sign.
Since distinct locations $y$ use disjoint sets of entries of the $A_j$, the answers at these remaining locations are independent uniform signs, independent of $B$ and the answers at collision locations.}
Therefore, 
\[
\Pr[\mathrm{success}]
\le 
\frac{1}{2} \operatorname{Pr}[\neg C]+\operatorname{Pr}[C]=
\frac12+\frac12\Pr[C]
\le
\frac12+\frac{(T/m)^m}{2N^{m-1}}.
\]
Achieving success probability at least $2/3$ requires
\begin{equation}
\label{eq:even-basic-lower-bound}
T\ge m\,3^{-1/m}N^{(m-1)/m}
=\Omega(m N^{1-1/m}).
\end{equation}

We now strengthen this bound to capture the dependence on $\varepsilon$. For $\sqrt{3/N}\le \eps \le 1/8$,\footnote{This choice ensures the probability in \eqref{eq:condition B} is smaller than 1, and the lower bound in \eqref{eq:tv bound} is positive.}
conditional on $\sigma$, we choose $B(y)$ independently according to
\be \label{eq:condition B}
\Pr[B(y)=b\mid\sigma]
=\frac{1+4\eps\sigma b}{2},
\qquad b\in\{\pm1\}.
\ee
Let $P_+$ and $P_-$ denote the joint distributions of the algorithm's randomness, the shifts, and all query answers when $\sigma=+1$ and $\sigma=-1$, respectively.
We next bound the total variation distance between $P_+$
and $P_-$.
Note that if $p$ and $q$ are distributions of $r$ independent
signs with means $4\eps$ and $-4\eps$, respectively, then
\[
\operatorname{TV}(p,q)
\le
\sqrt{1-\big(\sum\nolimits_z\sqrt{p(z)q(z)}\big)^2}
=
\sqrt{1-(1-16\eps^2)^r}
\le 4\eps\sqrt r,
\]
where the first inequality follows from Cauchy--Schwarz.
Averaging the bound and using \eqref{eq:even-revealing-count} gives
\begin{equation}
\label{eq:even-transcript-tv}
\operatorname{TV}(P_+,P_-)
\le 4\eps\,\mathbb E[\sqrt M]
\le 4\eps\sqrt{\mathbb E [M]}
\le 4\varepsilon
\sqrt{\frac{(T/m)^m}{N^{m-1}}}.
\end{equation}
On the other hand, conditional on $\sigma$,
\[
\mathbb E[T_{2m} \mid\sigma]=4\sigma\varepsilon,
\qquad
\operatorname{Var}(T_{2m}\mid\sigma)
=\frac{1-16\varepsilon^2}{N}\le\frac1N.
\]
Chebyshev's inequality gives
\[
\Pr[\sigma T_{2m}<2\varepsilon]
\le\frac{1}{4N\varepsilon^2}
\le\frac1{12}.
\]
Whenever the estimator succeeds and
$\sigma T_{2m}\ge2\varepsilon$, its output has sign $\sigma$.
It therefore determines $\sigma$ with probability at least $2/3-1/12=7/12$.
For equal prior probabilities, any test has success probability at most
$(1+\operatorname{TV}(P_+,P_-))/2$.
Hence \eqref{eq:even-transcript-tv} and $(1+\operatorname{TV}(P_+,P_-))/2\ge 7/12$ gives
\be
\frac16
\le\operatorname{TV}(P_+,P_-)
\le4\varepsilon
\sqrt{\frac{(T/m)^m}{N^{m-1}}},
\label{eq:tv bound}
\ee
which implies
\begin{equation}
\label{eq:even-precision-lower-bound}
T\ge
m\left(\frac{N^{m-1}}{576\varepsilon^2}\right)^{1/m}
\ge
\frac{m}{24}N^{1-1/m}\varepsilon^{-2/m},
\end{equation}
where the last inequality uses $m\ge2$.

The bound~\eqref{eq:even-basic-lower-bound}
already covers $\varepsilon\ge1/8$ and bounded $N$.
For sufficiently large $N$, applying
\eqref{eq:even-precision-lower-bound}  gives
$T=\Omega(mN)$ for all $\varepsilon<\sqrt{3/N}$.
Altogether, we obtain
\[
T=\Omega\left(
m\min\left\{N,N^{1-1/m}\varepsilon^{-2/m}\right\}
\right),
\]
as claimed.
\end{proof}

\begin{cor}
\label{cor:even-lower-bound-constant}
    For every fixed integer $k=2m\ge 4$ and constant accuracy $0<\varepsilon<1$, we have
    \be
    R_{\varepsilon}^{\na}(T_{k})
    =\Omega(m N^{1-1/m})=\Omega_k(N^{1-2/k}).
    \ee
\end{cor}

\begin{rmk}[The role of nonadaptivity]
\label{rmk:even-adaptivity}
The hard distribution above is not difficult for adaptive algorithms.
The collision estimate \eqref{eq:even-revealing-count}
uses the fact that the query sets are independent of
the hidden shifts.
An adaptive algorithm can recover $a_1,\ldots,a_{m-1}$
by querying the first $m-1$ character oracles, using $(m-1)n$ queries, hence recover the hidden shifts, since $t_j=a_1+\cdots+a_j$.
For any chosen $y$, it can then query $f_{2m}(y)$ and $A_j(y+t_j)$ using $m$ further queries.
These answers recover $\sigma$ for
Construction~\ref{construction:even-hard}, or $B(y)$ for the modified construction.
Thus Theorem~\ref{thm:even-lower-bound} is specifically a nonadaptive lower bound, it does not imply the same lower bound for unrestricted adaptive classical algorithms.
\end{rmk}

\subsection{Odd-cycle case}
\label{subsec:odd case}

We now extend the even-cycle lower bound to the odd-cycle case, which requires an additional algebraic ingredient. 
Since our even-cycle hard instances satisfy $W=\sigma I$ in Proposition \ref{prop:even-cancellation}, simply appending a factor $D_h H^{\otimes n}$ yields a normalized trace of magnitude at most $N^{-1/2}$. Thus, such a direct extension cannot preserve a constant trace gap. To overcome this obstruction, we introduce a three-Hadamard quadratic cap, which allows us to construct the odd-cycle hard distribution from the even-cycle one while preserving a constant trace gap. Since the additional phase functions are explicitly known, the same collision analysis applies.

Denote the quadratic phase function
\[
Q(x)=(-1)^{\sum_{1\le i<j\le n}x_ix_j},
\]
and define
\[
\eta =
\begin{cases}
\mathbf{1}^n,&n\equiv 1\pmod 4\\
0,&\text{otherwise}
\end{cases},
\]
where $\mathbf{1}^n=(1,\ldots,1)\in \mathbb{F}_2^n$.
Let
\[
c_n=
\frac1N\operatorname{Tr}
\left[
D_QH^{\otimes n} D_QH^{\otimes n}  D_{Q\chi_\eta}H^{\otimes n} 
\right].
\]

\begin{lem}
\label{lem:value of cn}
We have 
\[
|c_n|=
\begin{cases}
1/2,&n\text{ even}\\
1/\sqrt2,&n\text{ odd}
\end{cases}.
\]
\end{lem}

\begin{proof}
Write
\[
q(x)=\sum_{i<j}x_ix_j\pmod 2,\qquad
p(x)=\sum_i x_i\pmod 2,
\]
and let $u=1$ if $n\equiv 1\pmod 4$, and $u=0$
otherwise. Thus $Q(x)=(-1)^{q(x)}$ and
$\chi_{\eta}(x)=(-1)^{up(x)}$.
Expanding the normalized trace gives
\[
c_n=N^{-5/2}\sum_{a,b,c\in\mathbb F_2^n}
(-1)^{q(a)+q(b)+q(c)
+a\cdot b+b\cdot c+c\cdot a+up(c)}.
\]
Let $s=a+b+c$. Since $q(x+y)=q(x)+q(y)+p(x)p(y)+x\cdot y
\pmod 2,$ we have
\[
q(a)+q(b)+q(c)+a\cdot b+b\cdot c+c\cdot a
= q(s)+p(a)p(b)+p(b)p(c)+p(c)p(a)
\pmod 2.
\]
This choice of $q$ cancels the inner product terms $a\cdot b+b\cdot c+c\cdot a$ from the Hadamard matrices. Replacing $c=a+b+s$ gives
\beas
c_n &=& N^{-5/2}\sum_{a,b,s\in\mathbb F_2^n}
(-1)^{q(s) + u p(s) + p(a) p(b) + (p(s)+1+u) (p(a)+p(b)) } \\
&=& N^{-5/2} 
\sum_{s\in\mathbb F_2^n} (-1)^{ q(s) + u p(s) } 
\sum_{a,b\in\mathbb F_2^n}
(-1)^{ p(a) p(b) + (p(s)+1+u) (p(a)+p(b)) } .
\eeas
For each $A\in \{0,1\}$, $p(x)=A \pmod 2$ has $N/2$ solutions. Thus
\beas
c_n 
&=& N^{-5/2} 
\sum_{s\in\mathbb F_2^n} (-1)^{ q(s) + u p(s) } 
\frac{N^2}{4}
\sum_{A,B\in \{0,1\} }
(-1)^{ A B + (p(s)+1+u) (A+B) } \\
&=& \frac{1}{2\sqrt{N}}
\sum_{s\in\mathbb F_2^n} (-1)^{ q(s) + u p(s) } 
(-1)^{p(s)+1+u} \\
&=& \frac{(-1)^{1+u}}{2\sqrt{N}}
\sum_{s\in\mathbb F_2^n} (-1)^{ q(s) + (1+u) p(s) } .
\eeas
Finally, since $q(s)=\binom{|s|}{2}\pmod 2$, $p(s)=|s|\pmod 2$, and
\[
(-1)^{\binom{|s|}{2}+|s|}
=\operatorname{Re}[(1+i)i^{|s|}], \quad
(-1)^{\binom{|s|}{2}}
=\operatorname{Re}[(1-i)i^{|s|} ],
\]
we have
\[
\begin{aligned}
\sum_{s\in\mathbb F_2^n} (-1)^{q(s)+p(s)}
&=\operatorname{Re}\left[
(1+i)\sum_{w=0}^n\binom nw i^w
\right]
=\operatorname{Re}[(1+i)^{n+1}]
= \sqrt{2N} \cos \frac{(n+1)\pi}{4}
,\\
\sum_{s\in\mathbb F_2^n} (-1)^{q(s)}
&=\operatorname{Re}[(1-i)(1+i)^n]
= \sqrt{2N} \cos \frac{(n-1)\pi}{4} .
\end{aligned}
\]
So when $n\not\equiv 1\pmod 4$, we have $u=0$ and
\be
\label{eq:cn u=0}
c_n = - \frac{1}{2\sqrt{N}} \sum_{s\in\mathbb F_2^n} (-1)^{q(s)+p(s)}
= - \frac{1}{\sqrt{2}} \cos \frac{(n+1)\pi}{4}.
\ee
When $n\equiv 1\pmod 4$, we have $u=1$ and
\be
\label{eq:cn u=1}
c_n =  \frac{1}{2\sqrt{N}} \sum_{s\in\mathbb F_2^n} (-1)^{q(s)}
=  \frac{1}{\sqrt{2}} \cos \frac{(n-1)\pi}{4}.
\ee
The claims now follow directly.
\end{proof}

The choice of $\eta$ is essential when
$n\equiv 1\pmod 4$. Without the correction $\chi_{\eta}$, there would be no need to introduce $u$, as we could simply take $u=0$ throughout. Hence $c_n=0$ by \eqref{eq:cn u=0} when $n\equiv 1\pmod 4$.
By choosing $\eta={\bf 1}^n$, we instead have
$u=1$, which gives \eqref{eq:cn u=1}, and so $c_n \neq 0$. Thus the correction prevents the cancellation
and ensures that $|c_n|\ge 1/2$ for all $n$.

\begin{thm}[Odd-cycle nonadaptive lower bound]
\label{thm:odd-lower}
For every integer $m\ge2$ and $0<\varepsilon<1/2$,
\[
R_{\varepsilon}^{\na}(T_{2m+1})
=
\Omega\left(
m\min\left\{N,N^{1-1/m}\varepsilon^{-2/m}\right\}
\right).
\]
\end{thm}

\begin{proof}
We construct a hard distribution for $T_{2m+1}$ using Construction~\ref{construction:even-hard}, with the modification
introduced in the proof of Theorem~\ref{thm:even-lower-bound}.
The even-cycle hard distribution consists
of the $2m$ Boolean functions
\[
\begin{aligned}
g_{2j-1}(x)&=\chi_{a_j}(x),
&&1\le j\le m,\\
g_{2j}(x)&=A_j(x),
&&1\le j\le m-1,\\
g_{2m}(x)&=B(x)\prod_{j=1}^{m-1}A_j(x+t_j).
\end{aligned}
\]
We now define $2m+1$ Boolean functions by
\[
\begin{aligned}
f_1(x)&=Q(x)g_1(x),\\
f_j(x)&=g_j(x),
&&2\le j\le 2m-1,\\
f_{2m}(x)&=Q(x)g_{2m}(x),\\
f_{2m+1}(x)&=Q(x)\chi_{\eta}(x).
\end{aligned}
\]
Similarly to Proposition \ref{prop:even-cancellation}, $D_{g_1}H^{\otimes n}\cdots D_{g_{2m}}H^{\otimes n}=H^{\otimes n}D_BH^{\otimes n}.$
Then we have
\[
\begin{aligned}
D_{f_1}H^{\otimes n}\cdots D_{f_{2m+1}}H^{\otimes n}
&=
D_Q
\left(
D_{g_1}H^{\otimes n}\cdots H^{\otimes n}D_{g_{2m}}
\right)
D_QH^{\otimes n}D_{Q\chi_{\eta}}H^{\otimes n}\\
&=
D_QH^{\otimes n}D_BD_QH^{\otimes n}
D_{Q\chi_{\eta}}H^{\otimes n}.
\end{aligned}
\]
Consequently,
\begin{equation}
\label{eq:odd-weighted-mean}
T_{2m+1}
=
\frac1N\sum_y w_y B(y),
\end{equation}
where $w_y$ is the $y$-th diagonal entry
\[
w_y=
\left(
D_QH^{\otimes n}D_{Q\chi_{\eta}}H^{\otimes n}
D_QH^{\otimes n}
\right)_{yy}.
\]
Note that we have
\begin{equation}
\label{eq:odd-weights}
|w_y|\le1,
\qquad
\frac1N\sum_y w_y=c_n.
\end{equation}

We next prove the query lower bound.
Each query to $f_j$ for $j\le2m$ can be simulated by one
query to $g_j$ at the same location, followed by multiplication
by a known sign. The last oracle is completely known.
Thus the simulation preserves nonadaptivity and does not
increase the number of queries.
Consider a randomized nonadaptive algorithm making at most
$T$ queries, and let $C$ denote the collision event in the even-case:
\[
C=\left\{\,\exists
(x_1,\ldots,x_{m-1},y)
\in Q_2\times\cdots\times Q_{2m}:
x_j=y+t_j,\quad 1\le j\le m-1
\right\},
\]
The same collision counting arguments apply.
Taking $B\equiv\sigma$ gives $T_{2m+1}=\sigma c_n$.
Since $\varepsilon<1/2\le|c_n|$, any additive-$\varepsilon$
estimator with $\varepsilon<1/2$ can determine
$\sigma$ from the sign of its estimate divided
by $c_n$. Now the argument leading to
\eqref{eq:even-basic-lower-bound} therefore gives
\begin{equation}
\label{eq:odd-basic-lower-bound}
T\ge m\,3^{-1/m}N^{1-1/m}=\Omega(mN^{1-1/m}).
\end{equation}

For $\sqrt{3/N}\le\varepsilon\le1/16$, conditional on
$\sigma$, choose the values $B(y)$ independently with
mean $8\sigma\varepsilon$ and variance $1-64\varepsilon^2$ using a similar construction to \eqref{eq:condition B}.
By \eqref{eq:odd-weighted-mean} and \eqref{eq:odd-weights},
\[
\mathbb E[T_{2m+1}\mid\sigma]
=
8\sigma\varepsilon c_n,
\qquad
\operatorname{Var}(T_{2m+1}\mid\sigma)
=
\frac{1-64\varepsilon^2}{N^2}\sum_y w_y^2
\le\frac1N.
\]
Since $8\varepsilon|c_n|\ge4\varepsilon$, the same
concentration and testing argument as in the even-case proof
applies, using the sign of the estimate divided by $c_n$
to determine $\sigma$.
Together with the total variation estimate
with bias $8\varepsilon$, this gives
\begin{equation}
\label{eq:odd-precision-lower-bound}
T=\Omega\left(mN^{1-1/m}\varepsilon^{-2/m}\right).
\end{equation}
The remaining ranges of $N$ and $\varepsilon$ follow from
\eqref{eq:odd-basic-lower-bound} and monotonicity in
$\varepsilon$, as in the even-case proof, which yields the claimed complexity.
\end{proof}

\begin{cor}
\label{cor:odd-lower-bound-constant}
    For every fixed integer $k=2m+1\ge 5$ and constant accuracy $0<\varepsilon<1/2$, we have
    \be
    R_{\varepsilon}^{\na}(T_{k})
    =\Omega(m N^{1-1/m})=\Omega_k(N^{1-2/(k-1)}).
    \ee
\end{cor}

\section{A randomized nonadaptive algorithm}
\label{sec:even-upper}

For even $k$, we now present a randomized nonadaptive algorithm that estimates $T_k$ to additive error $\varepsilon$, with query complexity matching the lower bound in Theorem~\ref{thm:even-lower-bound}.

\begin{thm} 
\label{thm:even-upper}
For every even integer $k\ge4$,
$0<\varepsilon<1$, and $0<\delta<1/2$,
there exists a randomized nonadaptive classical
algorithm that estimates $T_{k}$ to additive
error $\varepsilon$ with success probability
at least $1-\delta$, using
\be
O\left(
k\min\left\{
N,\,
N^{1-2/k}\varepsilon^{-4/k}\log(1/\delta)
\right\}
\right)
\ee
oracle queries.
\end{thm}

\begin{proof}
Writing $x_{k+1}=x_1$, define
\bes
h(x_1,\ldots,x_k)
=
\bigg(\prod_{j=1}^k f_j(x_j)\bigg)
(-1)^{\sum_{j=1}^k x_j\cdot x_{j+1}}.
\ees
Expanding the normalized trace gives
\bes
T_k
=N^{k/2-1}\mathbb E[h(X_1,\ldots,X_k)],
\ees
where $X_1,\ldots,X_k$ are independent and
uniformly distributed over $\mathbb{F}_2^n$.

For each function $f_j$, independently sample
$q$ uniform points
$X_{j,1},\ldots,X_{j,q}\in \mathbb{F}_2^n$
and query $f_j$ at these locations.
All query locations are chosen in advance.
We define the estimator
\be
\label{eq:even-cartesian-estimator}
\widetilde T_k
=
\frac{N^{k/2-1}}{q^k}
\sum_{i_1,\ldots,i_k=1}^q
h(X_{1,i_1},\ldots,X_{k,i_k}).
\ee
Clearly, $\mathbb E[\widetilde T_k]=T_k.$
Thus, it remains to bound the variance.
For any nonempty subset $S\subseteq[k]$, define $\overline{S} = [k] \backslash S$ and
\bes
h_S(x_S)
:=
\mathbb E_{X_{\overline{S} }}
[h(x_S,X_{\overline{S} })],
\ees
where the coordinates in $S$ are fixed and
all remaining coordinates are independent and uniform.
We regard $[k]$ as the vertex set of the $k$-cycle graph.
A subset $S$ is called {\em independent} if it contains
no two adjacent vertices.

\begin{claim}
\label{claim:even-conditional-moments}
For every nonempty $S\subseteq[k]$, let $r=|S|$.
Then
\bes
\|h_S\|_2^2
\le
\begin{cases}
N^{r-k+1},
& S\text{ is independent},\\
N^{r-k},
& S\text{ is not independent}.
\end{cases}
\ees
In fact, equality holds in the second case.
\end{claim} 

\begin{proof}[Proof of Claim \ref{claim:even-conditional-moments}]
List the vertices of $S$ in cyclic order as
$v_1,\ldots,v_r$, with $v_{r+1}=v_1$.
For each $j$, consider the segment of the cycle
from $v_j$ to $v_{j+1}$, and denote its vertices
in order by $w_0=v_j,w_1,\ldots,w_{\ell}=v_{j+1}.$
Define $U_j=H^{\otimes n}D_{f_{w_1}}H^{\otimes n}
\cdots D_{f_{w_{\ell-1}}}H^{\otimes n}.$
Thus, the element
$U_j(x_{v_j},x_{v_{j+1}})$ represents the
sum over all intermediate variables along
this segment. In particular, if the two
vertices are adjacent, then $U_j=H^{\otimes n}$.
Expanding the matrix products gives
\bes
h_S(x_S)
=
N^{r-k/2}
\left(\prod_{v\in S}f_v(x_v)\right)
\prod_{j=1}^r
U_j(x_{v_j},x_{v_{j+1}}).
\ees
Define matrix $P_j$ via $P_j(x,y)=|U_j(x,y)|^2.$
Since $U_j$ is orthogonal, $P_j$ is doubly stochastic.
Taking the squared $L_2$ norm yields
\bes
\label{eq:even-moment-stochastic-trace}
\|h_S\|_2^2
=
N^{r-k}\operatorname{Tr}(P_1\cdots P_r).
\ees
Since a product of doubly stochastic matrices
is doubly stochastic,
$\operatorname{Tr}(P_1\cdots P_r)\le N.$
This proves the first bound.

If $S$ contains two adjacent vertices,
then at least one $U_j=H^{\otimes n}$, so $P_j=J/N,$
where $J$ is the all-ones matrix.
Since $PJ=JP=J$ for every
doubly stochastic matrix $P$, we obtain
$P_1\cdots P_r=J/N.$
Consequently, $\operatorname{Tr}(P_1\cdots P_r)=1,$
and hence $\|h_S\|_2^2=N^{r-k}.$ This proves the second bound.
\end{proof}

We now use Claim \ref{claim:even-conditional-moments} to
analyze the variance of the estimator.
Consider two terms in sum
\eqref{eq:even-cartesian-estimator}.
Let $S\subseteq[k]$ be the set of coordinates
in which their sample indices coincide.
This is the standard variance decomposition for (multi-sample) $U$-statistics~\cite{Hoeffding1948}.
Conditioned on the shared samples, all other
samples are independent. Hence the expectation
of the product of the two terms is
$\|h_S\|_2^2$.
For a fixed overlap pattern $S$, the probability
that two uniformly chosen index tuples coincide
exactly in the coordinates of $S$ is $q^{-|S|}\left(1-1/q\right)^{k-|S|}.$
Writing $\mu=\mathbb E[h]$, we therefore obtain
\bes
\begin{aligned}
\operatorname{Var}(\widetilde T_k)
&=
N^{k-2}
\sum_{\varnothing\neq S\subseteq[k]}
\frac{(1-1/q)^{k-|S|}}{q^{|S|}}
\left(\|h_S\|_2^2-\mu^2\right)\\
&\le
N^{k-2}
\sum_{\varnothing\neq S\subseteq[k]}
\frac{\|h_S\|_2^2}{q^{|S|}}.
\end{aligned}
\ees
Applying Claim \ref{claim:even-conditional-moments}
gives
\bes
\label{eq:even-variance-main}
\operatorname{Var}(\widetilde T_k)
\le
\sum_{\substack{\varnothing\neq S\subseteq[k]\\
S\text{ independent}}}
\frac{N^{|S|-1}}{q^{|S|}}
+
\sum_{\substack{\varnothing\neq S\subseteq[k]\\
S\text{ not independent}}}
\frac{N^{|S|-2}}{q^{|S|}}.
\ees
To bound the variance, we use the fact that every independent set $S$ of the even $k$-cycle has size at most $k/2$.
Choose
\[
q=\left\lceil
16N^{1-2/k}\varepsilon^{-4/k}
\right\rceil.
\]
If $q\ge N$, we query all $kN$ oracle values
and compute $T_k$ exactly.
Otherwise, we use the estimator above.
Let $M={N^{k/2-1}}/{q^{k/2}}.$
For an independent set $S$ with $r=|S|\le k/2$,
\[
\frac{N^{r-1}}{q^r}
=
\frac1N\left(\frac Nq\right)^r
\le M.
\]
For a non-independent set $S$, we have
$r=|S|\le k$, and similarly,
\[
\frac{N^{r-2}}{q^r}
=
\frac1{N^2}\left(\frac Nq\right)^r
\le M^2.
\]
Consequently, $\operatorname{Var}(\widetilde T_k)
\le 2^k(M+M^2).$
Our choice of $q$ implies $M \le\varepsilon^2/4^k$.
Since $k\ge4$ and $\varepsilon<1$,
\[
\operatorname{Var}(\widetilde T_k)
\le
{\varepsilon^2}/{2^k}
+
{\varepsilon^4}/{8^k}
<
{\varepsilon^2}/{4}.
\]
Chebyshev's inequality now gives
\[
\Pr\left[
|\widetilde T_k-T_k|>\varepsilon
\right]\le\frac14.
\]
Thus, the estimator achieves additive error
$\varepsilon$ with constant success probability, using
\[
O\left(
k\min\left\{
N,N^{1-2/k}\varepsilon^{-4/k}
\right\}
\right)
\]
queries.
To increase the success probability to $1-\delta$, we run the algorithm independently $O(\log(1/\delta))$ times and take the median of the estimates. This increases the query complexity by a factor of $O(\log(1/\delta))$.
\end{proof}

Although the above algorithm was presented for even $k$, its estimator and conditional-moment bound remain valid for odd $k$. 
We obtain the following upper bound.

\begin{cor}
\label{cor:odd-upper}
For every odd integer $k\ge5$ and constant $0<\varepsilon<1$, there exists a randomized
nonadaptive classical algorithm that estimates $T_k$ to additive error $\varepsilon$ with success probability at least $2/3$, using
\be
O_{k}(N^{1-2/k})
\ee
oracle queries.
\end{cor}

\begin{proof}
The estimator \eqref{eq:even-cartesian-estimator} and Claim~\ref{claim:even-conditional-moments} remain valid for odd $k$.
Thus, the same
variance bound applies.
Choose $q=\lceil C N^{1-2/k}\rceil$, where $C\ge1$ depends only on $k$ and $\varepsilon$.
Every independent set $S$ of the cycle graph satisfies
$|S|\le(k-1)/2$, so
\[
\frac{N^{|S|-1}}{q^{|S|}}
\le C^{-|S|}N^{2|S|/k-1}
\le C^{-|S|}.
\]
For every non-independent set $S$, we similarly have
${N^{|S|-2}}/{q^{|S|}}
\le C^{-|S|}$, since $|S|\le k$.
The same variance bound therefore gives
\[
\operatorname{Var}(\widetilde T_k)
\le
\sum_{\varnothing\ne S\subseteq[k]}C^{-|S|}
=
(1+C^{-1})^k-1.
\]
Choosing $C$ sufficiently
large and applying Chebyshev's inequality
completes the proof.
\end{proof}

\begin{rmk}
\label{rmk:odd-explicit}
The dependence on $\varepsilon$ can be made explicit.
For every odd $k\ge5$ and $0<\varepsilon<1/2$,
the same estimator achieves constant success probability
using
\[
O\left(
k\min\left\{
N,\,
\max\left\{
N^{1-2/(k-1)}\varepsilon^{-4/(k-1)},
N^{1-2/k}\varepsilon^{-2/k}
\right\}
\right\}
\right)
\]
oracle queries.
The first term inside the maximum dominates when $\varepsilon\le N^{-1/(k+1)}$.
In this regime, the upper bound matches the lower bound in Theorem~\ref{thm:odd-lower}.
For constant $\varepsilon$, a gap remains
between our lower and upper bounds:
\[
\Omega_k(N^{1-2/(k-1)})
\le R_\varepsilon^{\na}(T_k)
\le O_{k,\varepsilon}(N^{1-2/k}).
\]
\end{rmk}

\begin{rmk}
\label{rem:parity effect}
For odd $k\ge 5$, the variance analysis reveals
a specific obstacle to achieving the conjectured
query complexity $O_k(N^{1-2/(k-1)})$.
Indeed, suppose we choose
$q=CN^{1-2/(k-1)}$.
For every nonempty proper subset $S\subsetneq[k]$,
the variance contribution is bounded by
${N^{|S|-1}}/{q^{|S|}} \le C^{-|S|}$ if $S$ is independent, since $|S|\le (k-1)/2$; and by
$ {N^{|S|-2}}/{q^{|S|}} \le C^{-|S|}$, otherwise,
since $|S|\le k-1$.
The only obstruction comes from the full-overlap
term $S=[k]$, whose contribution to the variance
is exactly
\[
\frac{N^{k-2}}{q^k}(1-\mu^2)
=\Theta_{k,C}(N^{2/(k-1)}),
\]
where $\mu=\mathbb E[h]$ and
$|\mu|\le N^{1-k/2}$.
Thus, the present Cartesian-product estimator
cannot achieve bounded variance at the conjectured
query scale.
If this full-overlap contribution could be
suppressed while retaining the bounds for all
proper subsets, the resulting variance analysis would yield a matching nonadaptive upper bound.
Whether a different estimator can achieve this
remains an interesting open question.

\end{rmk}

\section{A fast randomized algorithm for estimating \texorpdfstring{$T_3$}{T3}}
\label{sec:three-cycle}

We show that $T_3$ can be estimated with
$O(n\log n)$ nonadaptive classical queries
for any fixed additive error and constant success probability.

Expanding the normalized trace gives
\bes
T_3=N^{-5/2}\sum_{a,b,c\in \mb F_2^n}
f_1(a)f_2(b)f_3(c)
(-1)^{a\cdot b+b\cdot c+c\cdot a}.
\ees
The key observation is the identity
\bes
(-1)^{a\cdot b+b\cdot c+c\cdot a}
=
i^{\mathrm{wt}(a)+\mathrm{wt}(b)+
\mathrm{wt}(c)-\mathrm{wt}(a+b+c)},
\ees
where $\mathrm{wt}(x)$ denotes the Hamming
weight of $x$. Define
\[
F_j(x)=f_j(x)i^{\mathrm{wt}(x)},
\qquad
\alpha_j(s)=\widehat F_j(s),
\quad j=1,2,3.
\]
Since $i^{-\mathrm{wt}(x)}
=\prod_{j=1}^n(-i)^{x_j}$, we have
\[
\widehat{i^{-\mathrm{wt}(\cdot)}}(s)
=\frac1{2^n}
\prod_{j=1}^n\left(1-i(-1)^{s_j}\right)
=\frac{(1-i)^n}{2^n}i^{\mathrm{wt}(s)}
=N^{-1/2}e^{-i\pi n/4}i^{\mathrm{wt}(s)}.
\]
Fourier inversion gives
\be
T_3=e^{-i\pi n/4}
\sum_{s\in \mathbb{F}_2^n }
i^{\mathrm{wt}(s)}
\alpha_1(s)\alpha_2(s)\alpha_3(s).
\ee

The Fourier representation shows that it suffices to identify the heavy Fourier coefficients of
$F_1$. Indeed, for any $\tau>0$, define $S_\tau=\{s\in \mathbb{F}_2^n:|\alpha_1(s)|\ge\tau\}.$
By Parseval's identity, $|S_\tau|\le\tau^{-2}$.
Moreover, Cauchy--Schwarz gives
\bes
\sum_{s\notin S_\tau}
|\alpha_1(s)\alpha_2(s)\alpha_3(s)|
\le
\tau\sum_{s\in\mathbb{F}_2^n } |\alpha_2(s)\alpha_3(s)|
\le
\tau\|\alpha_2\|_2\|\alpha_3\|_2
=\tau.
\ees
Thus, the total contribution from the terms indexed by $s\notin S_\tau$ is at most $\tau$ in absolute value.
We use the Goldreich--Levin algorithm~\cite[Section~3.5]{ODonnell2014} to identify the indices $s$ for which $|\alpha_1(s)|\ge\tau$. 
For completeness, we briefly explain how its Fourier-prefix search can be implemented with nonadaptive oracle queries in our setting.

For a prefix $p\in\mathbb F_2^\ell$, let
\[
W_\ell(p)
=
\sum_{t \in \mathbb{F}_2^{n-\ell} }|\alpha_1(p,t)|^2.
\]
Writing $x=(u,z)$, where
$u\in\mathbb F_2^\ell$ and
$z\in\mathbb F_2^{n-\ell}$, 
Parseval's identity gives
\be
\label{eq:three-prefix}
W_\ell(p)
=
\mathbb E_{ \substack{ u,v\in \mathbb{F}_2^\ell \\
z\in \mathbb F_2^{n-\ell}} }
\operatorname{Re}\left[
F_1(u,z)\overline{F_1(v,z)}
(-1)^{p\cdot(u+v)}
\right].
\ee
This is \cite[Proposition 3.40]{ODonnell2014}, adapted to complex-valued Boolean functions.

Each sample in \eqref{eq:three-prefix} requires only two queries to $f_1$, at locations $(u,z)$ and $(v,z)$. The key observation is that, for each fixed $\ell$, these query locations are independent of the prefix $p$, only the factor $(-1)^{p\cdot(u+v)}$ depends on $p$. Thus, the same query answers can be used to estimate $W_\ell(p)$ for different prefixes $p$. We can therefore make all oracle queries in advance and then identify the large Fourier coefficients using only classical postprocessing. In particular, the algorithm is nonadaptive.

\begin{prop}
\label{prop:three-upper}
For every $0<\varepsilon\le1$ and
$0<\delta<1$, there exists a randomized
nonadaptive classical algorithm that estimates
$T_3$ to additive error $\varepsilon$
with success probability at least $1-\delta$,
using
\be
\label{eq:three-query-bound}
O\left(
n\varepsilon^{-4}
\log\frac{n}{\delta\varepsilon}
\right)
\ee
oracle queries. In particular, for constants
$\varepsilon,\delta$, the query complexity is
$O(n\log n)$.
\end{prop}

\begin{proof}
Let $\tau=\Theta(\varepsilon)$ and
$M=\Theta(\varepsilon^{-2})$.
By the standard Goldreich--Levin algorithm, applied to the
prefix energies in \eqref{eq:three-prefix},
we can identify a set $S\subseteq\mathbb F_2^n$
such that
$S_\tau\subseteq S, |S|\le M,$
with probability at least $1-\delta/2$,
using
$O (
n\tau^{-4}\log(n/\delta\tau))
=O (
n\varepsilon^{-4}\log(n/\delta\tau))$
queries to $f_1$.
Here the prefix energies are estimated to
additive error $\Theta(\tau^2)$, and at most
$O(\tau^{-2})$ prefixes are examined at each
level. As observed above, all queries can be
chosen in advance, independently of the
prefixes examined during the search.

Next, let $\xi=\Theta(\varepsilon)$.
Independently query each oracle $f_j$ at $O(\xi^{-2}\log({M}/{\delta}))=O(\varepsilon^{-2} \log(1/\delta \varepsilon))$ uniformly random locations, and use these samples to estimate $\alpha_j(s)$ for every $s\in S$.
By Hoeffding's inequality and a union bound,
with probability at least $1-\delta/2$,
all these estimates satisfy
$|\widetilde\alpha_j(s)-\alpha_j(s)|
\le\xi.$
We output the estimate
\[
\widetilde T_3
=\operatorname{Re}\left[
e^{-i\pi n/4}
\sum_{s\in S}i^{\mathrm{wt}(s)}
\prod_{j=1}^3\widetilde\alpha_j(s)
\right].
\]
On the joint success event, the contribution
from indices outside $S$ is at most $\tau$.
Moreover, by the triangle inequality and Parseval's identity, we have
\[
\|\widetilde\alpha_j|_S\|_2
\le \|\alpha_j|_S\|_2
+\|(\widetilde\alpha_j-\alpha_j)|_S\|_2
\le
\|\alpha_j|_S\|_2+\xi\sqrt{|S|}
\le1+\xi\sqrt M
\le O(1).
\]
A telescoping expansion, followed by
Cauchy--Schwarz, gives
\[
\sum_{s\in S}
\bigg|
\prod_{j=1}^3\widetilde\alpha_j(s)
-
\prod_{j=1}^3\alpha_j(s)
\bigg|
\le
\xi\sum_{s\in S}
\Big(
|\alpha_2(s)\alpha_3(s)|
+
|\widetilde\alpha_1(s)\alpha_3(s)|
+
|\widetilde\alpha_1(s)\widetilde\alpha_2(s)|
\Big) 
\le O(\xi).
\]
Consequently, $|\widetilde T_3-T_3|
\le \tau+O(\xi) = O(\varepsilon).$
The success probability is at least
$1-\delta$.
The total number of queries is
$O (
n\varepsilon^{-4}\log(n/\delta\tau))$.
All oracle queries are chosen in advance,
so the algorithm is nonadaptive.
\end{proof}

\section{Adaptive classical query complexity lower bounds}
\label{sec:adaptive-lower-bound}

We briefly consider adaptive classical algorithms. 
By adapting Shepherd's hidden-subspace construction~\cite[Proposition 3.3.6]{Shepherd2009} to Boolean phase oracles, we obtain a uniform randomized query lower bound for cyclic Forrelation.

\begin{thm}
\label{thm:adaptive-lower-bound}
For every integer $k\ge4$ and $0<\varepsilon<1/2$,
\be
R_\varepsilon^{\ad}(T_k)=\Omega(N^{1/4}).
\ee
\end{thm}

\begin{proof}
We first consider $k=4$.
Let $L\le\F_2^n$ be a uniformly random subspace of
dimension $\lfloor n/2\rfloor$.
Choose independent uniform $s,t\in\F_2^n$ and
$\sigma\in\{-1,1\}$, and independent uniform Boolean
functions
$\phi:\F_2^n/L\to\{-1,1\}$ and
$\psi:\F_2^n/L^\perp\to\{-1,1\}$.
Define
\[
\begin{aligned}
f_1(x)&=\chi_s(x)\phi(x+L),&
\quad f_3(x)&=\chi_s(x)\phi(x+t+L),\\
f_2(x)&=\chi_t(x)\psi(x+L^\perp),&
f_4(x)&=\sigma\chi_s(t)\chi_t(x)\psi(x+s+L^\perp).
\end{aligned}
\]
Note that $\operatorname{supp}(\widehat f_1)\subseteq s+L^\perp$ since $\phi$ is constant on cosets of $L$.
Moreover, we have $\widehat f_3(a)=\chi_s(t)\chi_a(t)\widehat f_1(a)$ and $\mathbb E_x[f_2(x)f_4(x+a)]
=\sigma\chi_s(t)\chi_t(a)$ for $a\in s+L^\perp$.
Thus by the Fourier representation of $T_4$ in \eqref{eq: Fourier expression of T4},
\[
T_4
=\sum_{a\in\mathbb F_2^n}\widehat f_1(a)\widehat f_3(a)
\mathbb E_x[f_2(x)f_4(x+a)]
=\sigma\sum_{a\in s+L^\perp}
\widehat f_1(a)^2
\chi_s(t)^2\chi_a(t)\chi_t(a)
=\sigma\sum_{a\in s+L^\perp}\widehat f_1(a)^2
=\sigma.
\]

Consider an adaptive algorithm making at most $q$ queries.
We may assume that no oracle-location pair is queried twice.
Run the algorithm in an ideal experiment with four independent uniform random Boolean oracles.
The resulting adaptive query sequence is independent of the hidden parameters $L,s,t,\sigma$.

We couple this ideal experiment with the planted distribution. Suppose
the algorithm queries \(f_1\) at \(x\) with the answer $\chi_s(x)\phi(x+L).$
If the label \(x+L\) has not appeared previously, then
\(\phi(x+L)\) is still an unused independent uniform sign. We may
therefore choose its value so that the planted answer agrees with the
corresponding ideal answer. The same is true for queries to \(f_3\),
whose hidden \(\phi\)-label is \(x+t+L\), and similarly for \(f_2\)
and \(f_4\), whose hidden \(\psi\)-labels are \(x+L^\perp\) and
\(x+s+L^\perp\), respectively.
Consequently, the planted and ideal executions can be made identical
until the algorithm queries a hidden \(\phi\)- or \(\psi\)-label that
has already appeared before. Thus it suffices to bound the probability
that such a collision occurs.

Let $C$ denote such a collision along the ideal query
sequence.
Since this sequence is independent of $L,s,t$,
we may fix it when bounding $\Pr[C]$.
For two distinct queries to the same oracle, the
collision probability is
$(|L|-1)/(N-1)$ or $(|L^\perp|-1)/(N-1)$.
For queries to different oracles sharing $\phi$ or $\psi$,
the uniform shift gives collision probability
$|L|/N$ or $|L^\perp|/N$.
Since there are at most $\binom q2$
pairs of queries, the union bound gives
\[
\Pr[C]
\le\binom q2\frac{\max\{|L|,|L^\perp|\}}{N}
=O(q^2/\sqrt N).
\]
The ideal transcript and $C$ are jointly independent
of $\sigma$, and the two executions coincide on $\neg C$.
Consequently, any guess $\widehat\sigma$ satisfies
\[
\Pr[\widehat\sigma=\sigma]
\le\frac12\Pr[\neg C]+\Pr[C]
\le\frac12+O(q^2/\sqrt N).
\]
Since $T_4=\sigma$, a successful estimate with
$\varepsilon<1$ determines $\sigma$.
Success probability at least $2/3$ thus requires
$q=\Omega(N^{1/4})$.

For even $k>4$, append $k-4$ constant-one oracles and use $H^2=I$.
For $k=5$, let $g_1,\ldots,g_4$ be an input above.
Its four-oracle operator is orthogonal with normalized
trace $\sigma$, and hence equals $\sigma I$.
Using $Q,\eta,c_n$ from Subsection~\ref{subsec:odd case}, let
\[
(f_1,f_2,f_3,f_4,f_5)
=(Qg_1,g_2,g_3,Qg_4,Q\chi_{\eta}).
\]
The construction gives
$T_5=\sigma c_n$, where $|c_n|\ge1/2$.
Each query to this input requires at most one query
to the corresponding $g_j$, and $f_5$ is known.
A successful estimate with $\varepsilon<1/2$
therefore determines $\sigma$ from the sign of its
output divided by $c_n$.
The same lower bound follows for $T_5$.
Appending $k-5$ constant-one oracles extends it to
every odd $k\ge5$.
\end{proof}

\section{Conclusion}

In this work, we studied the classical query complexity of cyclic trace Forrelation. We obtained tight nonadaptive bounds for every fixed even cycle length, polynomial lower and upper bounds with a remaining exponent gap for odd cycle lengths, a polylogarithmic-query classical algorithm for \(T_3\), and a uniform polynomial lower bound for adaptive classical algorithms.
These results show that cyclic Forrelation exhibits a rich dependence on the cycle length and on whether classical queries are adaptive. Determining the optimal odd-cycle and adaptive complexities, and understanding how these phenomena extend to more general inter-query unitaries, remain interesting directions for future work.

\section*{Acknowledgment}

CS was supported by the National Key Research Project of China under Grant No. 2025YFA1017200. YZ was supported by the European Research Council (ERC) via the Starting Grant QuanThermal (101222179).
We would like to thank Kewen Wu for his helpful discussions.

\paragraph{AI disclosure:} The lower bound results in this paper were developed in part through interactive discussions with GPT-5.6 Plus, which was used to explore possible proof ideas and formulations. 
The manuscript was written by the authors, with assistance from GPT-6 in revising the exposition, polishing the language, and checking the arguments.
All mathematical statements and proofs were verified by the authors, who take full responsibility for the content of the paper.

\bibliographystyle{alphaurl}
\bibliography{dqc1.bib}

\end{document}